\documentstyle[aps,prl,epsfig]{revtex}

\begin{document}

\draft

\widetext

\title{ AGING IN THE RANDOM ENERGY MODEL}


\author{ G\'erard Ben Arous}
\address{DMA, EPFL, CH-1015 Lausanne, 
Switzerland}
\author{Anton Bovier}

\address{Weierstrass-Institut f\"ur Angewandte Analysis und Stochastik,
Mohrenstrasse 39, D-10117 Berlin, Germany.}
\author{ V\'eronique Gayrard}
\address{DMA, EPFL, CH-1015 Lausanne, 
Switzerland, and CPT-CNRS, Luminy, Case907, F-13288 Marseille Cedex 9, France}

\date{\today}

\date{\today}

\twocolumn[\hsize\textwidth\columnwidth\hsize\csname@twocolumnfalse\endcsname
\maketitle

\begin{abstract}
 In this letter we announce  rigorous results on 
 the phenomenon of aging 
in the Glauber dynamics of the random energy model and their relation to 
Bouchaud's 'REM-like' trap model. We show that, below the critical temperature,
 if we consider a time-scale that diverges 
with the system size in such a way that equilibrium is almost, but not quite
reached on that scale, a suitably defined autocorrelation function has the same asymptotic behaviour 
than its analog in the trap model.  
\end{abstract}


]

\narrowtext

\def\a{\alpha}
\def\b{\beta}
\def\d{\delta}
\def\e{\epsilon}
\def\ve{\varepsilon}
\def\f{\phi}
\def\g{\gamma}
\def\k{\kappa}
\def\l{\lambda}
\def\r{\rho}
\def\s{\sigma}
\def\t{\tau}
\def\th{\theta}
\def\vt{\vartheta}
\def\vp{\varphi}
\def\z{\zeta}
\def\o{\omega}
\def\D{\Delta}
\def\L{\Lambda}
\def\G{\Gamma}
\def\O{\Omega}
\def\S{\Sigma}
\def\Th{\Theta}
\def\del #1{\frac{\partial^{#1}}{\partial\l^{#1}}}

\def\h{\eta}

\def\E{{I\kern-.25em{E}}}
\def\N{{I\kern-.22em{N}}}
\def\M{{I\kern-.22em{M}}}
\def\R{{I\kern-.22em{R}}}
\def\Z{{Z\kern-.5em{Z}}}
\def\P{{I\kern-.22em{P}}}
\def\one{{1\kern-.25em\hbox{I}}}
\def\eu{{1\kern-.25em\roman{I}}}
\def\f1{{1\kern-.25em\roman{I}}}

\def\la{\langle}
\def\ra{\rangle} 
\def\der{\frac{d}{dx}}
\def\del{\partial}
\def\tr{\roman{tr}}
\def\dist{\,\roman{dist}}
\def\pb{\Phi_\b}
\def\AA{{\cal A}}
\def\BB{{\cal B}}
\def\CC{{\cal C}}
\def\DD{{\cal D}}
\def\EE{{\cal E}}
\def\FF{{\cal F}}
\def\GG{{\cal G}}
\def\HH{{\cal H}}
\def\II{{\cal I}}
\def\JJ{{\cal J}}
\def\KK{{\cal K}}
\def\LL{{\cal L}}
\def\MM{{\cal M}}
\def\NN{{\cal N}}
\def\OO{{\cal O}}
\def\PP{{\cal P}}
\def\QQ{{\cal Q}}
\def\RR{{\cal R}}
\def\SS{{\cal S}}
\def\TT{{\cal T}}
\def\VV{{\cal V}}
\def\UU{{\cal U}}
\def\VV{{\cal V}}
\def\WW{{\cal W}}
\def\XX{{\cal X}}
\def\YY{{\cal Y}}
\def\ZZ{{\cal Z}}
\def\A{{\cal A}}

\def\chap #1#2{\line{\ch #1\hfill}\numsec=#2\numfor=1\numtheo=1}
\def\sign{\,\roman{sign}\,}
\def\un #1{\underline{#1}}
\def\ov #1{\overline{#1}}
\def\ba{{\backslash}}
\def\sb{{\subset}}
\def\wt{\widetilde}
\def\wh{\widehat}
\def\sp{{\supset}}
\def\rar{{\rightarrow}}
\def\lar{{\leftarrow}}
\def\em{{\emptyset}}
\def\inn{\roman{int}\,}
\def\sminn{\roman{int}\,}
\def\smcl{\roman{cl}\,}

\def\ov#1{\overline{#1}}
\def\jg{J_\g}
\def\limlaw{\buildrel \DD\over\rightarrow}
\def\eqlaw{\buildrel \DD\over =}

\def\note#1{\footnote{#1}}
\def\tag #1{\eqno{\hbox{\rm(#1)}}}
\def\frac#1#2{{#1\over #2}}
\def\sfrac#1#2{{\textstyle{#1\over #2}}}
\def\ssfrac#1#2{{\scriptstyle{#1\over#2}}}
\def\text#1{\quad{\hbox{#1}}\quad}
\def\newpage{\vfill\eject}
\def\proposition #1{\noindent{\thbf Proposition #1}}
\def\datei #1{\headline{{}\hfill{#1.tex}}}
\def\theo #1{\noindent{\thbf Theorem {#1} }}
\def\fact #1{\noindent{\thbf Fact #1: }}
\def\lemma #1{\noindent{\thbf Lemma {#1} }}
\def\definition #1{\noindent{\thbf Definition {#1} }}
\def\hypothesis #1{\noindent{\thbf Hypothesis #1 }}
\def\corollary #1{\noindent{\thbf Corollary #1 }}
\def\proof{{\noindent\pr Proof: }}
\def\proofof #1{{\noindent\pr Proof of #1: }}
\def\endproof{$\diamondsuit$}
\def\remark{\noindent{\bf Remark: }}
\def\lg{\buildrel {\textstyle <}\over  {>}}
\def\var{\hbox{var}}
\font\pr=cmbxsl10

\font\thbf=cmbxsl10 scaled\magstephalf

\font\bfit=cmbxti10
\font\ch=cmbx12
\font\ftn=cmr8
\font\smit=cmti8
\font\ftit=cmti8
\font\it=cmti10
\font\bf=cmbx10
\font\sm=cmr7
\font\vsm=cmr6

\def\vep{\varepsilon}

Dynamical properties of complex disordered systems such as spin glasses, 
have been at the center of interest among theoretical physicists working 
in statistical mechanics. A key concept that has emerged over the last years
is that of ``aging'': it is applied to systems whose dynamics are 
dominated by slow transients towards equilibrium. When discussing aging dynamics,
it is all important to specify the precise {\it time-scales } considered in relation 
to the large volume limit. In "glassy dynamics" (see \cite{[BCKM]} for an excellent review) 
one considers the infinite volume limit at fixed time $t$, and {\it then} analyses the
 ensuing dynamics as $t$ tends to infinity. The typical setting where such a program has 
been carried out
is Langevin dynamics of sperical mean field spin glasses, such as the $p$-spin SK model 
\cite{[CD]}. Note that even in this setting multiple and even infinitely many time-scales 
may be observed (e.g. in the SK-model or the $2+4$-spin spherical SK model 
\cite{[BCKM]}). In this context,
mathematically rigorous results have been obtained recently in \cite{[BDG]} only for the $p=2$ 
spherical SK-model. 

On a heuristic level, one expects that glassy dynamics describes the evolution of a system that is 
approaching  ``local equilibrium", or nearest local minima of the free energy.
 However, the 
standard picture  of the spin glass phase typically involves a highly
complex landscape of the free energy function exhibiting 
many nested valleys organized according to some hierarchical tree structure
(see e.g. \cite{[BiK],[FS]}). To such a picture corresponds the
 heuristic image of a stochastic dynamics that, on time-scales that diverge with the 
size of the system,
can be described as performing a sequence of ``jumps'' between
different valleys at random times those rates are governed by the depths 
of the valleys and the heights of connecting passes or saddle points. 
To capture these features which are obviously beyond the realm of glassy 
dynamics, Bouchaud and others \cite{[B],[BD],[BM],[BMR],[BCKM]}
have introduced an interesting ansatz. Their so-called 
``trap models"  are Markov jump processes 
on a state space that simply enumerates the valleys of the free energy 
landscape. While this picture is
intuitively  appealing, its derivation is based on knowledge obtained in much 
simpler contexts, such as diffusions in finite 
 dimensional potential landscapes
exhibiting a finite number of minima. In the systems one is interested in 
here, however, both  dimension and number of minima are 
infinite or asymptotically growing to infinity.

It is thus an important and interesting question to understand whether, how, 
and in which sense the long time dynamics of disordered systems such as
spin glasses can really be described by trap models, and in particular
 to elucidate the precise time-scales
to which these models refer. To answer this 
question requires,
of course, the study of the actual stochastic dynamics of the full system at diverging 
time-scales, which is, in general, a very hard problem.

In this note we report on the first rigorous results linking the long-time behaviour of 
Glauber dynamics to a trap model  in the context 
of the ``simplest spin glass'' \cite{[GM]} model, the random energy model
(REM) 
\cite{[D1],[D2]}.
While this model is surely far from ``realistic'', it offers a number of 
features that are ``typical'' for what one expects in real spin glasses,
and its analysis involves already a good number of the problems one would 
expect to find in more realistic situations. The main advantage we
will draw from this is, of course, the fact that the equilibrium properties of 
this model are perfectly understood.

\noindent {\bf The REM}. We recall that the REM \cite{[D1],[D2]}
 is defined as follows.
A spin configuration $\s$ is a vertex of the hypercube 
$\SS_N\equiv \{-1,1\}^N$. 
We define the family of i.i.d. standard Gaussian random variables 
$\{E_\s\}_{\s\in\SS_N}$.  We define a
 random  (Gibbs) probability measure on $\SS_N$, $\mu_{\b,N}$, 
by setting 
$$
\mu_{\b,N}(\s)\equiv \frac {e^{\b \sqrt N E_\s}}{Z_{\b,N}}
\label{1.1}
$$
where $Z_{\b,N}$ is the normalizing partition function.

It is well-known \cite{[D1],[D2]} that this model exhibits a phase transition 
at $\b_c=\sqrt{2\ln 2}$. For $\b\leq\b_c$, the Gibbs measure is supported, 
asymptotically as $N\uparrow \infty$, on the set of states $\s$ for which
$E_\s\sim \sqrt N\b$, and no single configuration has positive mass. For 
$\b>\b_c$, on the other hand, the Gibbs measure gives positive  mass 
to the extreme elements of the order statistics of the family $E_\s$;
i.e. if we order the spin configurations according to the magnitude of their  
energies s.t.
\begin{equation}
E_{\s^{(1)}}\geq E_{\s^{(2)}}\geq E_{\s^{(3)}}\geq\dots\geq E_{\s^{(2^N)}}
\label{1.2}
\end{equation}
then for any finite $k$, the respective mass $\mu_{\b,N}(\s^{(k)})$
will converge, as $N$ tends to infinity, to some positive 
random  variable $\nu_k$; in fact, the entire family of masses
$\mu_{\b,N}(\s^{(k)}), \k\in\N$ will converge to 
a random process $\{\nu_{k}\}_{k\in\N}$, called 
Ruelle's point process \cite{[Ru],[B]}.  

So far the fact that $\s$ are vertices of a hypercube has played no r\^ole
in our considerations. It will enter only in the definition of the
{\it dynamics of the model}. The dynamics we  consider is a 
{\it
Glauber dynamics} 
$\s(t)$ on  $\SS_N$ with 
transition rates 
$$
p_N(\s,\eta)=e^{-\b \sqrt N E_\s},
\label{1.3}
$$
when $\s$ and $\eta$ differ by a single spin flip.
Note that the dynamics is also random, i.e. the law of the Markov chain is
a measure valued random variable on $\O$ that takes values in the space of 
Markov measures on the path space $\SS_N^{\N}$. We will mostly take a 
quenched point of view, i.e. we consider the dynamics for a given fixed 
realization of the disorder.

It is easy to see that this dynamics is reversible with respect to the
Gibbs measure $\mu_{\b,N}$. On also sees that it represents a 
nearest neighbor random walk on the hypercube with traps of 
random depths.

\noindent {\bf The REM-like trap model.} 
The idea suggested by the known behavior of the equilibrium 
distribution is that this dynamics, for $\b>\b_c$, will spend long 
periods of 
time in the states $\s^{(1)},\s^{(2)},\dots$ etc. and will move ``quickly''
from one of these configurations to the next. Based on this intuition,
Bouchaud et al. \cite{[B],[BD]} 
 proposed the ``REM-like'' trap model: consider a continuous time
Markov process $Z_M$ whose state space is
  the set  $S_M\equiv\{1,\dots,M\}$ of
 $M$ points, representing the $M$ ``deepest'' traps. Each of the 
states is assigned a positive random 
energy $E_k$ which is taken to be exponentially
distributed with rate one. 
 If the process is in state $k$, it 
waits an exponentially distributed time with mean proportional 
to $e^{E_k\a}$, and then jumps
with equal probability in one of the other states $k'\in S_M$. 

\def\cosec{\,\hbox{\rm cosec}\,}
This process can be  analyzed using  techniques from {\it renewal}
theory. The  point is that if one starts the process from 
the uniform distribution, it is possible to show that  
 the counting process, $c(t)$, that counts the number of 
jumps in the time 
interval $(0,t]$, is a 
classical renewal (counting) process \cite{[KT]}; 
moreover, as $n\uparrow\infty$,
this renewal process converges to a renewal process with a
{\it deterministic} law for the renewal time with a heavy-tailed
distribution whose density is proportional to $t^{-1-1/\a}$  
where $\a=\b/\b_c$. 

The  quantity that is used to  characterize the ``aging'' phenomenon is
the  probability $\Pi_M(t,s)$ that during a time-interval $[t,t+s]$ the 
process does not jump.
 Bouchaud and Dean \cite{[BD]} showed that, for $\a>1$, 
$$
\lim_{M\uparrow\infty}
\frac {\Pi_M(t,s)}{H_0(s/t)} =1,
\label{B.8.1}
$$
where the function $H_0$ is defined by 
\begin{equation}
H_0(w)\equiv \frac 1{\pi\cosec(\pi/\a)} 
\int_{w}^{\infty} dx   \frac {1}{(1+x)x^{1/\a}}
\label{B.53}
\end{equation}
Note that $H_0(w)$ behaves like $1-C w^{{1-1/\a}}$, for small $w$, and
like $C w^{-1/\a}$ for large $w$.

Our purpose is to show, in a mathematically rigorous way, how and to what 
extent the REM-like trap model can be viewed as an approximation
of what happens in the REM itself.
To this end we first introduce the set $T_M\equiv\{\s^{(1)},\dots,\s^{(M)}
\}$ of the first $M$ states defined by the enumeration 
\ref{1.2}.
 Ideally, we would like to start with 
our original process $\s(t)$ and construct a new process $Y_M$ as follows.
Let $\t_1,\t_2,\dots,\t_n,\dots$ be the sequence of times at which $\s(t)$
visits different elements of the (random) set $T_M$. Then 
set
$$
X_{N,M}(t)\equiv \sum_{k=1}^M k\,\one_{\s^{(k)}=\s(\t_n)}\one_{\t_n\leq t<\t_{n+1}}
\label{B.56}
$$
i.e. $X(t)$ takes the value $k$ during time-intervals 
at which the process $\s(t)$ ``travels'' from  $\s^{(k)} $ to the next element 
of this set. 
We would like to say that Bouchaud's process $Z_M$ 
approximates, after an $N$ and $M$ dependent  rescaling of the time, this
process $X$, if $N$ and $M$ are large, i.e. that in some appropriate sense, for 
some function $c(N,M)$, 
$$
Z_M(t)\sim X_{N,M}(c(N,M)t)
$$
when first $N$, then $M$, and finally $t$ tend to infinity.
 This problem involves two main assumptions:

\noindent{1)} The process jumps  with the uniform distribution 
from any state in $T_M$ to any of the other  states in $T_M$, and

\noindent{2)} The process observed on the set $T_M$  is, at least asymptotically,  a Markov process, in particular,
the times between visits
of two different states in $T_M$ are asymptotically exponentially distributed. 

\smallskip
\noindent
While it  appears intuitively reasonable to accept these assumptions, they are   a) not at all easy to justify  and b)
the second assumption is not even  correct. 
In fact, we will see  that such properties 
can be only established in a very weak asymptotic form, which is, however,
just enough to imply that the predictions of Bouchaud's model apply to the 
long time asymptotics of the process.

We will now present our main results. The full 
proofs of these results will be given 
in two forthcoming papers \cite{[BBG1],[BBG2]}.

We define instead of the set $T_M$ introduced above  the sets, 
for $E\in\R$,
$$
T\equiv T_{N}(E)\equiv \left\{\s\in \SS_N\big |E_\s\geq u_N(E)\right\}
\label{B.59}
$$
where
$$
u_N(x)\equiv \sqrt{2N\ln 2} +\frac x{\sqrt{2N \ln 2}} -
\frac 12\frac {\ln(N\ln2)+\ln 4\pi}{\sqrt {2N\ln 2}}
\label{2.8}
$$
We will call this set the ``top''. Note that $T_N(E)=T_{|T_N(E)|}$.

Our first result concerns just the ``motion'' of the process
disregarding time. 
Let $\xi^1,\dots,\xi^{|T(E)|}$ be an enumeration of the elements of $T(E)$.
Now define (for fixed $N$ and $E$), the stochastic  process $X_\ell$ with 
state space $\{1,\dots,|T(E)|\}$ and discrete time $\ell\in \N$ by
\begin{equation}
Y^{(N)}_\ell=X(\t_\ell)
\label{M1.5}
\end{equation}
It is easy to see that $Y^{(N)}_\ell$ is a Markov process whose transition probabilities $p(i,j)$ are 
nothing but the probabilities that the original process $\s(t)$ starting at $\xi^i$ hits $T$ first in
the point $\xi^j$.
To formulate our first theorem it will be convenient to fix the random size of the state space of this 
process  by
 conditioning. Thus set $P_M(\cdot)\equiv P(\cdot|\,|T(E)|=M)$. 

\theo{1} {\it Let $\s(n)$ denote the Markov chain with transition matrix
defined in (1.3) and  whose         initial distribution is the uniform 
distribution on $\SS_N$.
Let $Y^{(N)}_\ell$ be the Markov process defined by 
\ref{M1.5}. 
Let $Y_\ell$ denote the Markov chain on $\{1,\dots,M\}$ with transition 
matrix $p^*_M$ given by
$$
p^*_M(i,j)=\cases{ \frac 1{M-1},&\,{if}\,\,\, $i\neq j$\cr
 0,&\,{if}\,\,\,$ i= j$}
\label{M1.7}
$$
and initial distribution $p^*_M(i)=1/M$. 
 Then, for all $M\in \N$,
$$
Y^{(N)}\limlaw Y,\text{$P_M$-a.s.}
\label{M1.8}
$$
}

\remark Note that the statement of the theorem 
 implies the convergence in law (w.r.t. $P$) of 
the probability distribution of $Y^{(N)}$.

The next results concern the mean times of the motion towards the top and 
between  points of the top.

\theo{2.} {\it  Assume that $\a\equiv\b/\sqrt {2\ln 2}>1$. Then
on a set of probability one for all large enough $N$,
 the
following holds:
\begin{itemize}
\item [(i)]  For all $\eta\in \SS_N$, let $\TT(\eta)$ denote the
mean time the process starting in $\eta$ takes to reach $T_N(E)\ba \eta$.  Then
$$
\TT(\eta)=
\frac{M}{M-1}
\left[e^{\b\sqrt{N}E_{\eta}}+\WW_{N,E}
\right](1+O(1/N))
\label{3.1}
$$
\item [(ii)] For all $\eta, \bar\eta\in T(E)$, 
the expected time, $Z(\eta,\bar\eta)$, 
to reach $\bar\eta$ starting from $\eta$ conditioned
on the event that $\bar\eta$ is the first site different from 
$\eta$  in $T_N(E)$ that
is reached, is almost independent of $\bar \eta$, more precisely
\begin{equation}
\left|Z(\eta,\bar\eta)-\TT(\eta)
\right|
\leq
\frac{1}{1-\sfrac{1}{M}}\WW_{N,E}O(1/N)
\label{3.1ter}
\end{equation}
where $\WW_{N,E}$ is a random variable of mean value
$$
\E \WW_{N,E}=\frac{e^{\b\sqrt{N}u_N(E)}}{\a-1}
\label{3.003bis}
$$
and whose standard deviation is negligible compared to the mean, as $E$ 
tends to $-\infty$. 
\item [(iii)] If $\a<1$, then, for all $\s\in\SS_N$,
$$
\TT(\s) = \frac M{M-1} e^{N(\b^2/2+\ln
2)}(1+O(1/N))
\label{3.003.1}
$$
\end{itemize}
}

\remark  
We see that for $-E$ very large,
$\WW_{N,E}\sim e^{\b\sqrt{N}u_N(E))}$ represents a natural time-scale
for the process on $T(E)$.
Thus (i) implies that for all $\s\not\in T(E)$, 
the mean time of arrival in the top is proportional to 
$e^{\b\sqrt{N}u_N(E)}$.
On the other hand, there exists $\eta\in T(E)$ such that 
the mean time of first exit from $\eta$ which is  
$\sqrt N E_\eta$, is just of this order. Thus the slowest times of exit from a state 
in $T(E)$ are of the same order as the time it takes to reach $T(E)$. 
This can be expressed by saying that  on the
average the process takes a time $t$ to reach states that have an exit
 time $t$. This is a 
first manifestation of the aging phenomenon. 
 In contrast, if $\a<1$, for all $\s\in T(E)$, 
$\E\t^\s_{T(E)\ba\s}\ll \sup_{\eta\in\SS_N}\E \t^\eta_{\SS_N\ba\eta}$,  
and thus the time spent in top states is irrelevant compared to the
time between successive visits of such states.
Thus we see a clear distinction between the high and the low
temperature phase of the REM on the dynamical level. 
Notice that, as has been observed in \cite{[FIKP]}, the dynamical phase transition is
not accompanied by a qualitative change in the spectral gap, which in all 
cases is related to the largest exit times. For related results on the high 
temperature dynamics see also \cite{[MP]}.

\remark Statement iii) of Theorem 2 expresses the fact that the mean 
times of passage from a state $\eta\in T(E)$ to another state 
$\bar\eta\in T(E)$ are asymptotically independent of the terminal state 
$\bar \eta$. This confirms to some extent the heuristic picture of Bouchaud.
Indeed, if in addition we added the hypothesis that the process observed
on the top is Markovian, then the two preceeding theorems would immediately
imply that the waiting times must be exponentially distributed with rates
independent of the terminal state and given by \ref{3.1ter}. This, however,
cannot be justified. The reason for the failure of the Markov properties can 
 be traced to the fact that the time spent outside of $T(E)$ when travelling between 
two states of $T(E)$ cannot be neglected in comparison to the waiting time in the starting 
point, which in turn is a manifestation of the {\it absence} of a true 
{\it separation of time-scales}.

We now turn to a more precise analysis of the aging phenomenon. 

 The natural generalization of Bouchaud's correlation function
$\Pi_M(t,s)$ is the probability that the process does not
jump from a state in the top to another state in the top during a time 
interval of the form $[n, n+m]$. There is
some ambiguity as to how this should be defined precisely, but the most
convenient definition is to define 
$
\Pi_\s(n,m)$ as the probability that the process starting at time $0$
in $\s$ does not jump during the time interval $[n,n+m]$ from one
state in $T$ to another state in $T$. 

Of course we still have to specify the initial distribution. To be as
close as possible to Bouchaud, the natural choice is the uniform
distribution on $T_N(E)$ that we will denote by $\pi_E$. The natural
correlation function is then 
$$
\Pi(n,m)\equiv \frac 1{|T_N(E)|}\sum_{\s\in T_N(E)}\Pi_\s(n,m)
\label{B.63}
$$
The following theorem establishes the connection to the trap model:

\theo{3.}{\it Let $\b>\sqrt {2\ln 2}$. Then 
there is a sequence  $c_{N,E}\sim \exp(\b \sqrt Nu_N(E))$ such
that for any $\e>0$
$$
\lim_{t,s\uparrow\infty}\lim_{E\downarrow
-\infty}\lim_{N\uparrow\infty}
P\left[\left|\frac
{\Pi(c_{N,E}t,c_{N,E}s])}{H_0(s/t)}-1\right|>\e\right]=0
\label{T.1}
$$
where $H_0 $ is defined in \ref{B.53}.}

\remark Note that the rescaling of the time by the factor 
$c_{N,E}$ ensures that we are observing the system first on the proper
time 
of equilibration as $N$ goes to infinity, and that then, 
as $E$ tends to 
minus infinity, we measure  time on a  scale at which the 
equilibration time diverges. Thus the trap model describes the large time
asymptotics on the ``last scale'' before equilibrium is reached. 
This is to be contrasted to the other extreme of ``glassy dynamics'', when 
the infinite volume limit is taken with a fixed time scale.

The proof of this theorem is rather involved and is the main content of
\cite{[BBG2]}.
It may be instructive to see a brief outline of our  approach. 
Basically the
idea is to mimic the proof in the trap model and to set up
a renewal equation.
Now it is  easy to derive a renewal equation for the quantities 
$\Pi_\s(n,m)$.
However, in contrast to the situation of the trap model, it is not
possible
to obtain a single closed equation for $\Pi(m,n)$.
This means that we actually have to study  a system of renewal equations
 which renders the proof rather complicated. The
key  ingredients then are precise estimates of the Laplace transforms 
of the probability distributions entering the renewal equations in the complex plane.
What makes the final result emerge is then the fact that in the neighborhood
of the origin (which corresponds to large times), the Laplace transforms have 
almost the desired properties that 
would lead to such a closed equation. This makes it possible to employ perturbation 
expansions to prove the theorem. 

\noindent {\bf Acknowledgements:} We thank J.-Ph. Bouchaud, L. Cugliandolo,
and J. Kurchan for helpful  discussions and encouraging comments.

\end{document}